\documentclass[]{spie}  

\usepackage{amsmath,amsfonts,amssymb}
\usepackage{graphicx}
\usepackage[colorlinks=true, allcolors=blue]{hyperref}
\usepackage{multirow}
\usepackage{enumitem} 
\usepackage[]{algorithm2e}

\usepackage{xcolor}
\usepackage{subfigure}
\usepackage[normalem]{ulem}
\usepackage{mathtools}
\usepackage{amsmath}
\usepackage{ulem}
\usepackage{txfonts}
\usepackage{textcomp} 

\title{High contrast imaging wavefront sensor referencing from coronagraphic images}

\author[a,b,c]{Nour Skaf}
\author[b,d,e,f]{Olivier Guyon}
\author[a]{Anthony Boccaletti}
\author[b]{Vincent Deo}
\author[b,d]{Sébastien Vievard}
\author[b]{Julien Lozi}
\author[b]{Kyohoon Ahn}
\author[g]{Barnaby Norris}
\author[h]{Thayne Currie}
\author[a]{Eric Gendron}
\author[a]{Arielle Bertrou-Cantou}
\author[a]{Florian Ferreira}
\author[a]{Arnaud Sevin}
\author[a]{Fabrice Vidal}

\affil[a]{LESIA, Observatoire de Paris, Univ.PSL, CNRS, Sorbonne Univ., Univ.de Paris, 5 pl. Jules Janssen, 92195 Meudon, France}
\affil[b]{National Astronomical Observatory of Japan, Subaru Telescope, 650 North A'oh\=ok\=u Place, Hilo, HI 96720, U.S.A.}
\affil[c]{Department of Physics and Astronomy, University College London, London, United Kingdom}
\affil[d]{Astrobiology Center of NINS, 2-21-1 Osawa, Mitaka, Tokyo 181-8588, Japan}
\affil[e]{Steward Observatory, University of Arizona, Tucson, AZ 85721, USA}
\affil[f]{College of Optical Sciences, University of Arizona, Tucson, AZ 85721, USA}
\affil[g]{Sydney Institute for Astronomy, School of Physics, University of Sydney, NSW 200}
\affil[h]{NASA-Ames Research Center, Moffett Field, California, USA}

\authorinfo{Further author information: (Send correspondence to N.S)\\.: E-mail: nour.skaf@obspm.fr}

\begin{document} 
\maketitle

\begin{abstract}

A key challenge of high contrast imaging (HCI) is to differentiate a speckle from an exoplanet signal. The sources of speckles are a combination of atmospheric residuals and aberrations in the non-common path. Those non-common path aberrations (NCPA) are particularly challenging to compensate for as they are not directly measured, and because they include static, quasi-static and dynamic components. 
The proposed method directly addresses the challenge of compensating the NCPA. The algorithm DrWHO - Direct Reinforcement Wavefront Heuristic Optimisation - is a quasi-real-time compensation of static and dynamic NCPA for boosting image contrast. It is an image-based lucky imaging approach, aimed at finding and continuously updating the ideal reference of the wavefront sensor (WFS) that includes the NCPA, and updating this new reference to the WFS. Doing so changes the point of convergence of the AO loop. We show here the first results of a post-coronagraphic application of DrWHO. DrWHO does not rely on any model nor requires accurate wavefront sensor calibration, and is applicable to non-linear wavefront sensing situations. We present on-sky performances using a pyramid WFS sensor with the Subaru coronagraph extreme AO (SCExAO) instrument.

\end{abstract}


\section{INTRODUCTION}
\label{sec:intro}  

Adaptive Optics (AO) instrumentation has undergone extensive growth in sophistication and scientific capabilities over the last 30 years, enabling high-contrast imaging technologies, with extreme AO (ExAO) systems consisting of $\approx$2000-actuator DMs driven at over 1 kHz speeds \cite{GPIMacintosh2014,Beuzit2019}. 
This has lead to key scientific breakthroughs, 
including the first direct image of a planet-mass companion \cite{chauvin2004} and the first imaged system of jovian exoplanets, HR8799 \cite{Marois2008}.  

Some of the current generation of ExAO systems, e.g. the Subaru Coronagraphic Extreme Adaptive Optics (SCExAO) at the Subaru Telescope \cite{Jovanovic2015} and MagAO-X on the Magellan Clay telescope \cite{Males2018MagAOx} are pushing further achievable performance -- e.g. Strehl ratio (SR), 
contrast and sensitivity \cite{Vigan2015,Currie2020, VievardSPIE, DeoSPIE, Guyon2021SelfCalWFS}--  and serve as technology prototypes for ExAO instruments on 25-40 m extremely large telescopes
\cite{Kasper_PCS,TMTwhitepaper_Fitzgerald, GMT_2020, AhnSPIE}.\\

\begin{figure}[t]
\centering
\includegraphics[width=0.45\textwidth]{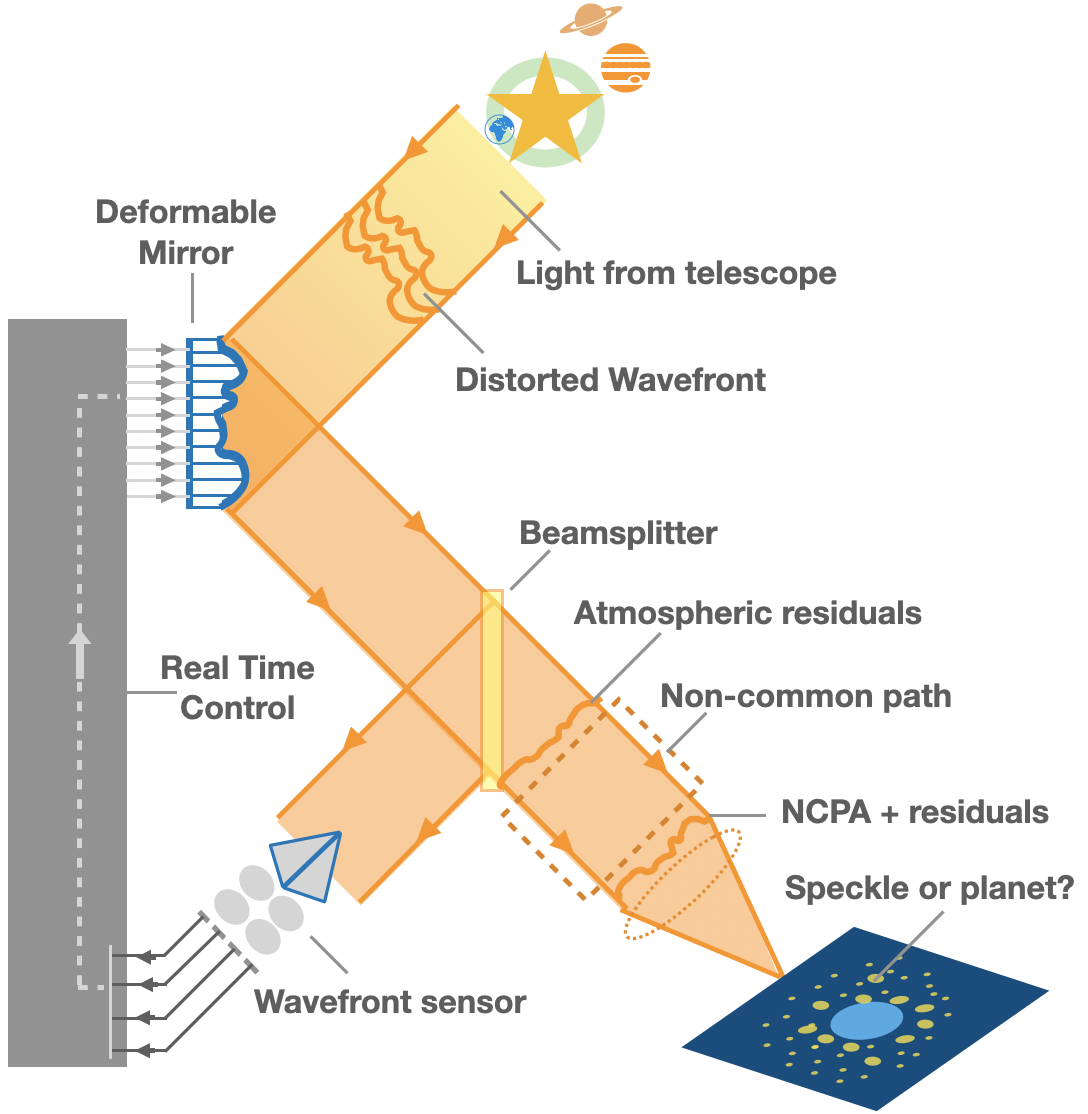}
\caption{Schematic view an AO system, presenting the problem of NCPA, especially in the case of exoplanet imaging with a coronagraphic image.}
\label{fig:AO_schema}
\end{figure}

The ExAO loop corrects wavefront aberrations measured by the WFS, ideally approaching an aberration-free image. However, the ExAO loop convergence point, defined by the WFS reference, may not correspond to the optimal image quality. Reasons for such a discrepancy include:
\begin{itemize}
    \item{\textbf{Non-common path aberrations} (NCPA) due to optics located after the beam splitting between WFS and science paths, inducing differential aberrations between the science camera and the WFS, as seen on Figure~\ref{fig:AO_schema}.
    In a high contrast imaging system, these aberrations may include coronagraph optics defects. These aberrations can also vary with temperature and mechanical deformation at timescales from minutes to hours, or with the positioning errors of moving optics, making them particularly challenging to calibrate. They are typically of the order of tens of nanometers, enough to lead to static and quasi static speckles in coronagraphic images \cite{sauvage2007, Vigan2019}. These speckles are also present in non coronagraphic images but less concerning because they do not contribute significantly to the SR, as usually beneath the camera contrast limit for unsaturated images}. 
    \item{\textbf{WFS calibration errors} due for example to a non-flat wavefront used to acquire a calibration.}
    \item{\textbf{Chromaticity} for systems where the wavelength of the WFS and the scientific camera are different.}
    \item{\textbf{Choice of optimal wavefront state}. The goal of the AO system may not be to produce a flat wavefront. For example, in a HCI, cancellation of static diffraction features (Airy diffraction rings, telescope spiders), such as dark hole techniques \cite{Potier2020darkhole}, may seek to drive the AO loop to a non-flat wavefront state.}
\end{itemize}

The phenomena responsible for the effects described above have timescales slower than the millisecond-level atmospheric turbulence timescale, and can be (nearly) static. 
In this paper, we refer to all discrepancies between WFS reference and optimal image wavefront as NCPA, noting that in previous publications it usually only refers to differential optical defects.

NCPA affect all AO systems, but are particularly acute in high contrast imaging where small wavefront errors can easily mask faint exoplanet images (cf Figure \ref{fig:AO_schema}). Furthermore, quasi-static aberrations do not average out to a spatially smooth halo as fast as the atmospheric speckles, leaving structures in the coronagraphic image which resemble planets, unlike speckles generated by atmospheric turbulence alone.


The most reliable approach to measure and correct optical misalignments is to use focal plane images~\cite{Gonsalves-div-82}. In the particular case of NCPA, these images are ideally generated by the science detector. 
Several solutions have been suggested for static NCPA \cite{Paxman1988, Frazin2018,Vigan2018NCPA, vievardSPIEWFSscexao, BosFF2020, Potier2020darkhole}, the quasi static part being particularly challenging  to measure and correct. The evolutionary timescale of the aberrations with respect to the frequency of the correction is a main difficulty.

In a closed loop AO system, NCPA are usually compensated by subtracting a biasing reference signal
in the WFS reference corresponding to the aberration to correct. Doing so requires good WFS response knowledge and stability, so that the adequate offset can be applied and maintained. Ideally, this offset should be updated in nearly real-time due to optical and mechanical variations in the instrument, however in practice it is typically updated on a daily or weekly basis. Finding the adequate offset requires focal plane images to update the reference, and can furthermore be particularly challenging if the WFS exhibits non-linear response that must be accounted for, as is the case in some high-performance WFSs such as the Pyramid WFS (PyWFS) \cite{Esposito2020,Deo2019gain,Chambouleyron2020, Chambouleyron2021}.

In this paper we will present the DrWHO algorithm, a model-free focal plane wavefront sensing approach which is aimed at finding the offset mentioned above to correct slow and static wavefront aberrations, including the NCPA, on a few seconds timescale. After stating the issues to be solved in Section \ref{sec:problem_statement}, we describe the algorithm in Section \ref{sec:algo_des}. The application of the algorithm to non-coronagraphic PSF, with numerical simulations and on-sky tests, is presented in Section \ref{sec:PSF_tests}.
Section \ref{sec:coro_tests} presents the main results of the on-sky validation for coronagraphic images. In Section \ref{sec:Discussion}, we review the characteristics of the algorithm and mention further implications for use in HCI.


\section{Problem statement: defining a good WFS reference}
\label{sec:problem_statement}
In a closed loop AO system, the WFS signal corresponds to the  WFS measurement subtracted by the WFS reference, to isolate residual wavefront errors that must be corrected by the DM. This reference defines the convergence point of the AO control loop, which brings the WFS signal to where it gives a flat wavefront after the real-time controller performs a wavefront reconstruction. 

This WFS reference is usually measured with an internal light source, before on-sky observations.
In reality, however, the \textit{ideal} WFS reference is different from the internal source WFS reference due to optical illumination discrepancies. Furthermore, the WFS reference contains a dynamic part, constantly evolving due to quasi-static aberrations and wavefront chromaticity. The accurate calibration of the WFS, through the reference, is therefore a significant challenge for any AO system, on top of the fact that the static part cannot be accurately measured.

Therefore, a continuous way to measure the \textit{ideal} reference is required for high accuracy wavefront correction. 

Finding this \textit{ideal} reference is especially critical to ExAO systems, where a slight deviation can introduce slow/static speckles that appear similar to a planet image.

\section{Algorithm description}
\label{sec:algo_des}
Because the \textit{ideal} reference changes continuously, the role of DrWHO is to update frequently, on seconds timescales, the \textit{actual} reference, and ensure the AO loop converges towards a correction of NCPA. In the context of high contrast imaging, the algorithm does so by wisely selecting the WFS images corresponding to the best lucky imaging of coronagraphic or non-coronagraphic images, in terms of contrast or intensity for the former, and Strehl ratio or other metric for the latter. Hence, the WFS and coronagraphic images first need to be synchronized in time. \\

The algorithm proceeds as follows: on a defined timescale $T_{DrWHO}$ (of the order of tens of seconds for example), the algorithm first selects the best frames according to a score metric, and a predefined selection fraction $P$, in percent. Then, out of those $P$\% frames, the corresponding WFS raw frames are extracted from the AO telemetry and averaged; the resulting WFS frame replaces the WFS \textit{actual} reference, and the algorithm is iterated to continuously optimize the WFS reference. \\
Hence, as DrWHO proceeds, the point of convergence of the AO loop includes NCPA and other aberrations of a timescale longer than $T_{DrWHO}$, which are thus compensated for. The image quality then improves and presents less speckles features. 
In fact, the frequency of the aberrations that DrWHO can correct depends on the timescale of one iteration of the algorithm, which depends whether there is a coronagraph in place or not, and how good the seeing conditions are.  

\begin{figure*}[t]
\begin{center}
\centering
\includegraphics[width=0.95\textwidth]{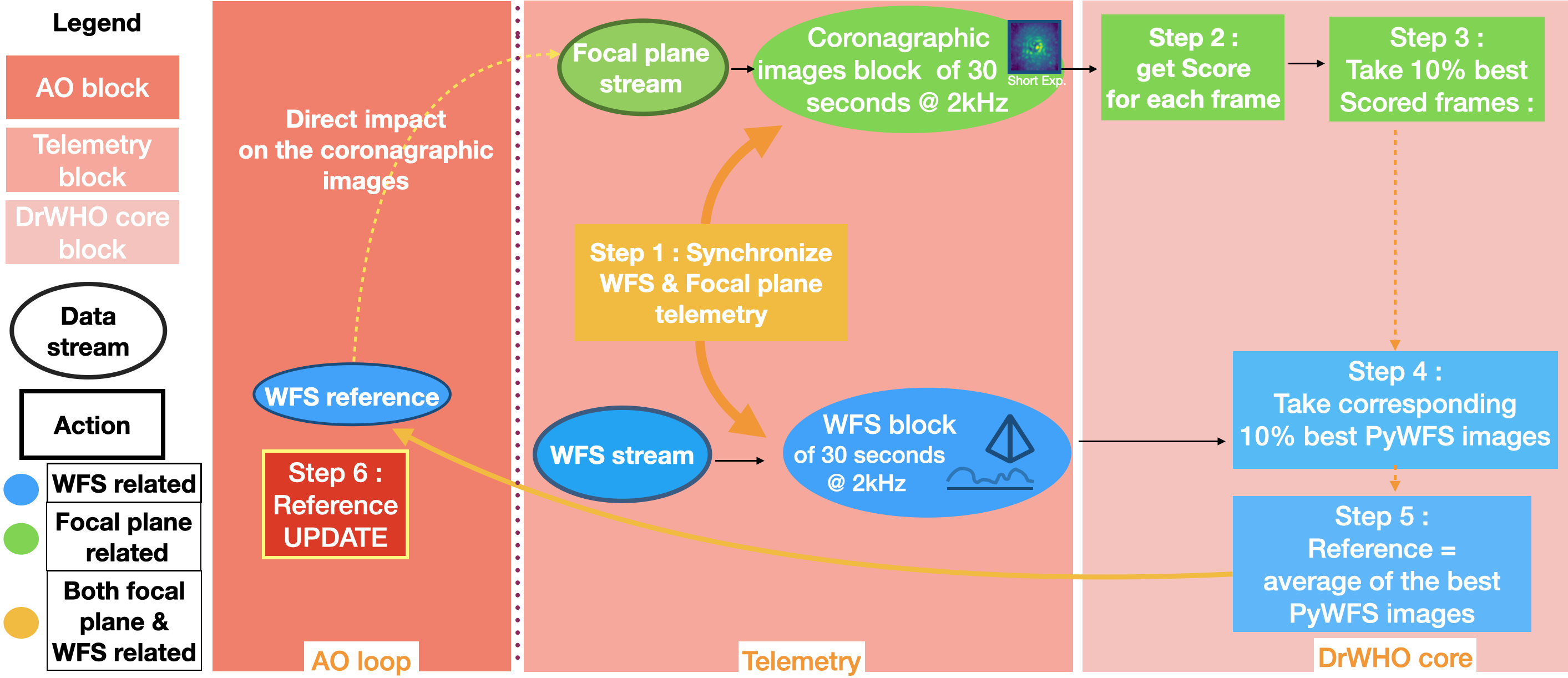}
\caption{Schematic view of DrWHO. From left to right, as the AO loop is running, the coronagraphic images, or focal plane images, and WFS telemetry are acquired. These data streams are then synchronised and re-sampled to the same timescale, here spanning the length of the DrWHO iteration, $T_{DrWHO}$ (s). On the PSF data cube, the second step of DrWHO is to compute the score for each frame. The following step is to select a fraction ($P$\%) of the best PSF according to the quality criteria. The 4th step is to take the corresponding WFS images, then (5th) to average them to make one single PyWFS image : this corresponds to the new reference for the iteration. Finally, the 6th and final step is to apply this new reference to the PyWFS, and thus update it. This last step is the only one not transparent to the AO loop, as it changes its point of convergence.}
\label{fig:DrWHOschema}
\end{center}
\end{figure*}

We present the procedure step by step of DrWHO with the  description in Algorithm \ref{algo:DrWHO}, and a schematic view of the algorithm is in Figure \ref{fig:DrWHOschema}.\\

\begin{algorithm}
\KwData{\begin{itemize}[noitemsep,nolistsep]
     \item WFS frames
     \item fast science camera frames
 \end{itemize}}
 \KwResult{Update the reference of the WFS}
 \textbf{Initialization:} take the reference of the WFS with the internal light source\;
 \While{AO loop is running}{
 \begin{itemize}
     \item acquire WFS and fast science focal plane coronagraphic images\;
     \item synchronize data cubes\;
    \While {DrWHO iteration ($T_{DrWHO}$ (seconds)) is running} {
    \begin{itemize}
        \item select the $P$\% best coronagraphic science data\;
        \item select the corresponding WFS images\;
        \item average selected WFS images\;
        \item the resulting WFS frame replaces the WFS reference\;
        \item WFS reference updated. 
    \end{itemize}
   }
   \item restart DrWHO iteration
   \end{itemize}
   Change of AO point of convergence;
  }
\caption{DrWHO algorithm}
\label{algo:DrWHO}
\end{algorithm}

\section{Non-coronagraphic results}
\label{sec:PSF_tests}
Most of those results have been extensively presented in \citenum{DrWHO1}, we present here a summary of those results. 

\subsection{Validation via numerical simulation}
\subsubsection{Simulation setup}
Numerical simulations have been performed with the AO simulator COMPASS \cite{Compass}. Compass is a versatile AO simulator, designed to meet the need of high performance for the simulation of AO systems, modeling several kinds of AO features, including several types of WFS, atmospheric simulations, DMs, telescopes, and RTCs. DrWHO was first implemented on the COMPASS software, to explore the algorithm feasibility and potential. Table \ref{tab:simuParams} synthetizes the parameters used in the COMPASS simulations.\\

\begin{table}[t]
		\centering
		\caption{%
			ExAO numerical simulations parameters.
		}
		\label{tab:simuParams}
		
		\renewcommand{\arraystretch}{1.2}
		\begin{tabular}{ll}
			\multicolumn{2}{c}{\textbf{Numerical simulation configuration}}\\
			\hline\hline
			\multirow{4}{*}{Telescope} & $D$ = 8.~m diameter\\
			& $\quad$No support spiders\\
			& Central Obstruction = 0.12 \\
			\hline
			\multirow{5}{*}{Turbulence layer} & von Kármán, ground layer only\\
			& $r_0$ = 0.16 at 500~nm \\ 
			& L$_0$ = 20~m \\
			& (to simulate post-woofer residuals) \\
			& $||\overrightarrow{\mathbf{v}}||$ = 10~m.s$^{-1}$\\
			\hline
			PyWFS & \\
			$\quad$Subapertures & 64$\times$64 \\
			$\quad$Wavelength & Monochromatic, 750~nm\\
			$\quad$Guide star magnitude & 7 \\
			$\quad$Modulation & Circular, 3~$\frac{\lambda}{D}$ radius, 24 points\\
			$\quad$Photon noise only \\
			\hline
			Deformable mirror & \\ 
			& 48 actuators across the diameter\\
			\hline
			Focal Plane Camera & \\
			$\quad$Wavelength & Monochromatic, 1.65~$\mu$m\\ 
			$\quad$Photon noise only \\
			\hline
			RTC controller & \\
			$\quad$Loop rate & 2~kHz\\
			$\quad$Method & Linear modal integrator\\
			$\quad$Basis & DM Karhunen-Loève (KL) basis \\
			$\quad$Loop gain & 0.3\\
			$\quad$Controlled modes & 1505\\
			$\quad$Modes filtered & 300\\
			\hline
		\end{tabular}
	\end{table}

The simulation is idealized on several aspects. First, it does not take into consideration the telescope spiders. Second, there is no source of noise other than only the photon noise on both the WFS and the science detector (observing mR=7 with 50\% efficiency). Furthermore, we considered a PyWFS with a higher sampling than on the SCExAO instrument (64 pixels over the PyWFS pupil diameter, versus 50 pixels), hence higher performance than achieved on-sky. The PyWFS is working in the visible at 750~nm, and the detector in the infrared at 1650~nm (cf Table \ref{tab:simuParams}).
Because COMPASS does not simulate coronagraphic images, DrWHO was first tested on non-coronagraphic PSF, with the Strehl ration as quality criteria for the image selection.

\subsubsection{Simulation results}
We performed a test with 15 DrWHO iterations of 10 000 iteration each (corresponding to 5 seconds at 2kHz). The number of AO loop per DrWHO iteration was constrained by simulation time. NCPA were applied on the science path, thus not visible by the WFS. Those NCPA correspond to a linear combination of the first 12 Karhunen-Loève modes (which is the linear and orthogonal modal basis used for computing the response matrix), with a randomized amplitude, and a total amplitude of 30~nm RMS, which is slightly greater to what is expected in reality (approximately 20~nm RMS on SCExAO).\\

Before presenting the results in terms of SR, it is important to emphasize that on most ExAO systems, the impact of NCPA over the SR is relatively small. 
Hence, it is not the best metric to quantify the efficiency of the algorithm, because the AO is dominated by dynamical wavefront residuals. 
In fact, NCPA are usually small enough that they do not affect the SR significantly relative to dynamical wavefront residuals. Their main contribution is to add quasi-static speckles in the focal plane, which mimic exoplanets signal. This is even more true when a coronagraph is used. However, the SR is a good metric to make sure the algorithm does not diverge, in particular through the high-order modes as they could adopt a random walk behavior. Furthermore, the simulation presents a simplified system where a coronagraph is not simulated, hence the SR remains useful.

\begin{figure}[t]
\begin{center}
\includegraphics[width=0.55\textwidth]{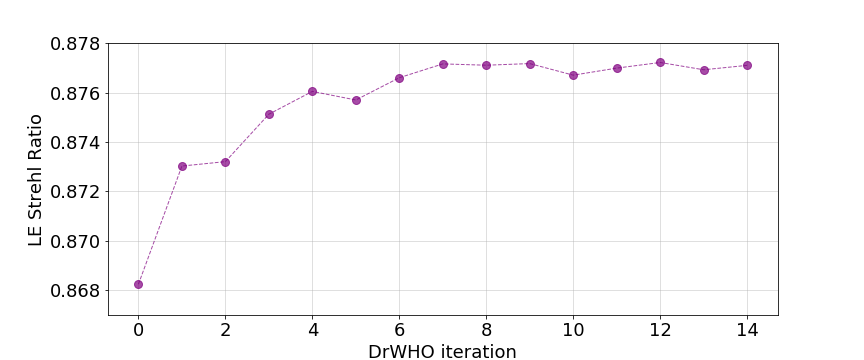}
\caption{Evolution of the long exposure (LE) Strehl ratio during the DrWHO run on COMPASS, corresponding to 15 DrWHO iterations of 10 000 AO loop iterations each.}
\label{fig:compass_LE_SR}
\end{center}
\end{figure}

Table \ref{tab:compass_comp} and Figure \ref{fig:compass_LE_SR} present the result of the SR after the DrWHO run: the final SR after the 15-iteration DrWHO run reaches 87.7\%, with a maximum of 87.9\% when there are no NCPA. 
As shown on Figure \ref{fig:compass_LE_SR}, nearly half of the SR improvement is obtained within the first iteration of DrWHO. Then, the SR keeps slowly improving, until it reaches a plateau. After 15 iterations, the SR reached is almost the same as the SR measured without NCPA. \\

\begin{table}[]
\centering
\caption{Comparison of the long exposure (LE) Strehl ratio between the following cases of the AO loop: no NCPA, NCPA alone, and post DrWHO with the NCPA. The second line corresponds to the RMS of the wavefront for each of those cases, calculated from the SR of the line above with the Maréchal approximation.}
\begin{tabular}{l|ccc}
                               & \multicolumn{1}{l}{No NCPA} & \multicolumn{1}{l}{with NCPA} & \multicolumn{1}{l}{with DrWHO} \\ \hline
SR LE (\%)                          & 87.9                      & 86.8                         & 87.7                           \\
$\sigma_{OPD}$ (nm RMS) & 94.3                      & 98.8                         & 95.1                         
\end{tabular}
\label{tab:compass_comp}
\end{table}

\begin{figure}[t]
\begin{center}
\includegraphics[width=0.55\textwidth]{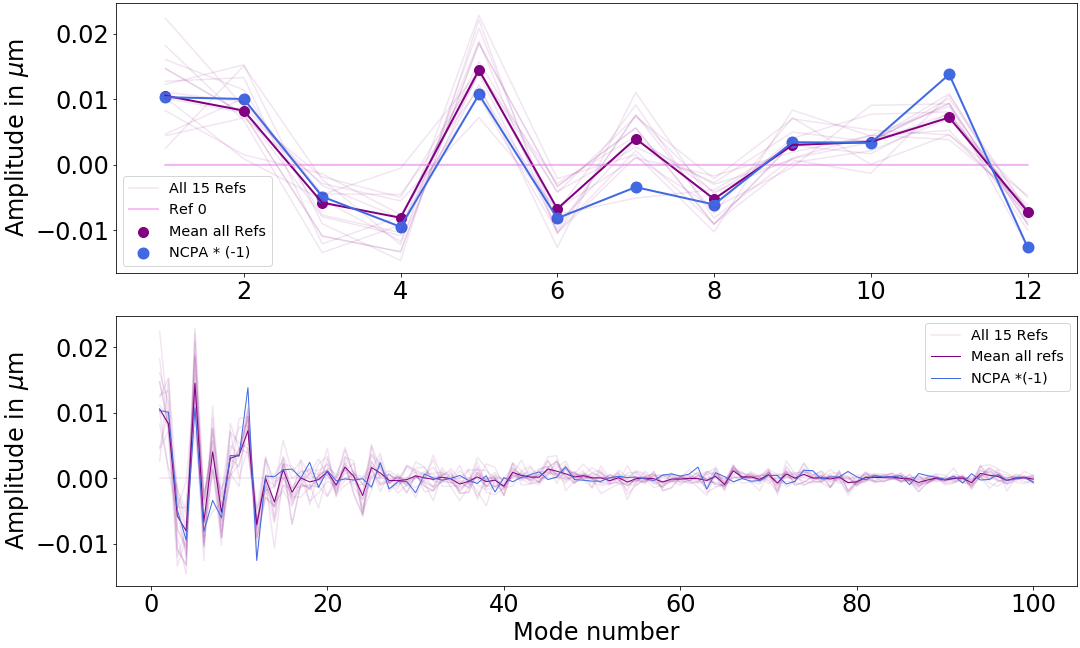}
\caption{Top: projection of the references provided by the 15-iteration DrWHO run over the modal basis, with the projection of the opposite of the NCPA, in order to better visualise the correction. Bottom: projection over the first 100 modes. The trend for the higher order modes is the same.}
\label{fig:ho_modes}
\end{center}
\end{figure}

In order to have a better understanding of the correction in terms of modes, the 15 references for each iteration of the DrWHO run were projected over the modal basis with which the response matrix was acquired. Figure \ref{fig:ho_modes} shows this modal decomposition along with the ideal NCPA correction, where the opposite of the NCPA was added, to better compare the correction. It appears from the top plot of this figure that, as we previously concluded, the algorithm converges from the first iteration towards a compensation of the NCPA. The average of all those references (minus the reference 0, which is the initial reference) is neighbouring the NCPA aberrations. This proves that DrWHO measured them, with the correct amplitude, and compensated for them. 
The middle and bottom plots of this figure prove that the algorithm does not diverge at the higher order modes, as confirmed by the increase of SR. 
The quadratic sum of 12 modes from the mean of all the references from the algorithm corresponds to \textbf{36~nm} RMS, which is roughly consistent with the amplitude of the NCPA.\\

To conclude the simulation results, DrWHO on COMPASS improved the image quality in terms of SR, and the NCPA applied over the PSF, were almost totally corrected. Half of the correction has occurred at the first iteration of the algorithm. The projection of the references from DrWHO is showing that the NCPA are nearly entirely compensated for, and the higher orders, where we did not apply NCPA, are not diverging.

\subsection{Validation with on-sky tests}

DrWHO was then adapted and deployed on the SCExAO instrument. This instrument is equipped with a Pyramid WFS operating in the 600-950~nm wavelength range \cite{Lozi2019PWFS}. The real time control (RTC) of the AO system is managed by the CACAO software - Compute And Control for Adaptive Optics - \cite{cacao2018}. The DM has 45 actuators across the pupil, giving a control radius of 22.5$\lambda$/D. We use the CACAO software to interact with the system and implement the algorithm, to communicate between the PyWFS and the DM.
Several scientific modules are downstream SCExAO. For testing DrWHO on non-coronagraphic images, the two VAMPIRES camera were used, operating in the visible \cite{vampires2015}.
The PyWFS and the VAMPIRES camera run at different frequencies, with typical frequencies of 2kHz for the former, and between 200 Hz to 2kHz for the latter, hence there is the need to synchronize the data streams to run DrWHO. 

\subsubsection{PSF quality criteria}
The Flux Concentration (FC) has been introduced in \citenum{DrWHO1} as the quality criteria for quantifying the results of the DrWHO run. It is defined as
\begin{equation}
\centering
    \text{FC} = \frac{\sum_i  (x_i^{\alpha})}{ (\sum_i x_i ) ^{\alpha}},
    \label{eq:FC}
\end{equation}

with $x_i$ being the value of the pixel i, and $\alpha > 1$ - we will use  $\alpha$ = 2. If some pixels have a negative value following a dark subtraction due to camera readout noise, then the pixel value is set to zero. 
This quantity is independent of absolute flux, since it is normalized by the total flux as seen in the denominator of equation \ref{eq:FC}. 

\begin{figure}[t]
\centering
\includegraphics[width=0.32\textwidth]{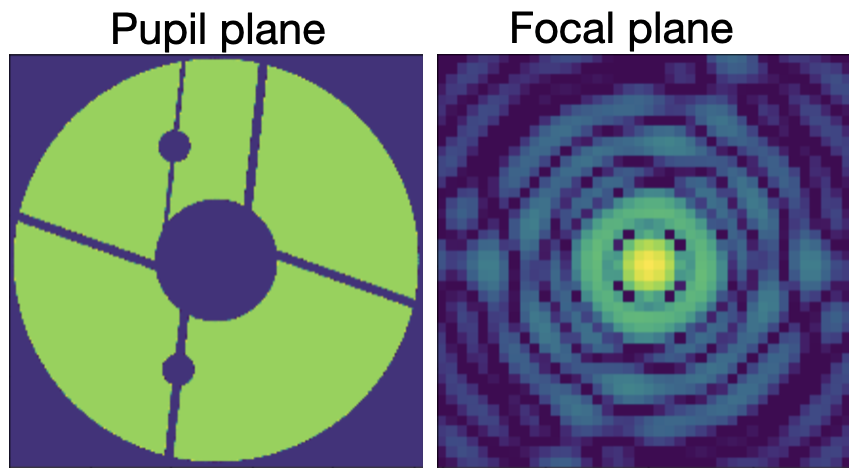}
\caption{Left: simulated pupil plane, including the spiders and a mask for two dead actuators. Right: simulated perfect SCExAO PSF
}
\label{fig:perfect_psf}
\end{figure}

The FC can be normalised by the FC of a simulated perfect PSF---which we will call 
FC$_0$ - 
computed for the pupil of the telescope (cf Figure \ref{fig:perfect_psf}), as the FC is a relative number. 
This new quantity is referred to as the normalized FC (nFC): nFC = FC / FC$_0$.

\subsubsection{Observation setup}

DrWHO was tested on-sky on the engineering night of December 8 UT, 2020. 20 iterations of the algorithm were performed, each of 30 seconds. Each iteration entails synchronized data cubes resampled to a common 1 kHz frame rate for both VAMPIRES and PyWFS images, running at 500Hz and 2kHz respectively. 1200 modes were corrected. The observation conditions at that time were a seeing of 0.5", and a wind speed of 4 m/s. The observed target was the star $\xi$ Leo, at an airmass of 1.03. The window size of the image for the selection of the quality criteria was of 70x70 pixels, corresponding to 0.43x0.43 arcsec field of view.

\subsubsection{PSF quality evolution}

\begin{figure*}[t]
\begin{center}
\centering
\includegraphics[width=0.8\textwidth]{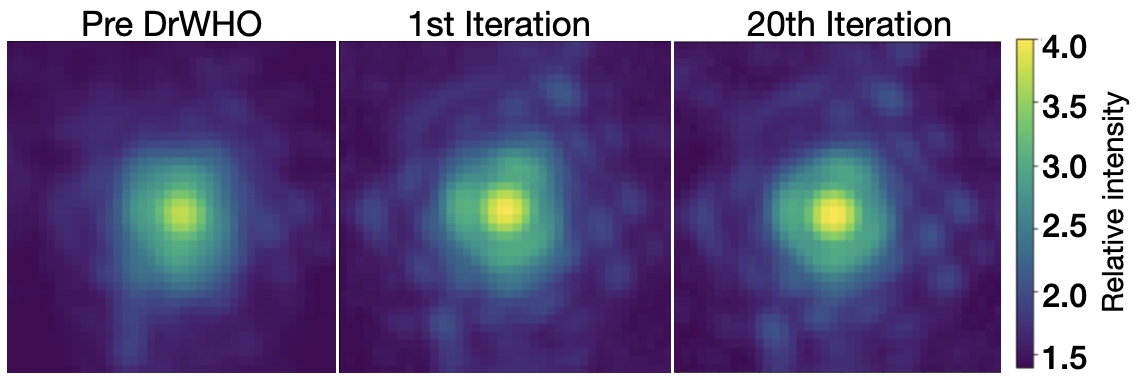}
\caption{Evolution of the on-sky PSF before running the algorithm, after the first iteration, and the after last iteration. Each image is 0.25~arcsec (40x40 pixels) across, acquired at $\lambda = $ 750~nm, 30 sec exposure time (computed by co-addition of 15,000 frames acquired at 500~Hz). Those PSFs should be compared to the ideal PSF in Figure \ref{fig:perfect_psf}, with matching wavelength, field of view and orientation of Figure \ref{fig:SkyResult} shown in logarithmic brightness scale.}
\label{fig:SkyResult}
\end{center}
\end{figure*}

\begin{figure}[t]
\centering
\includegraphics[width=.47\textwidth]{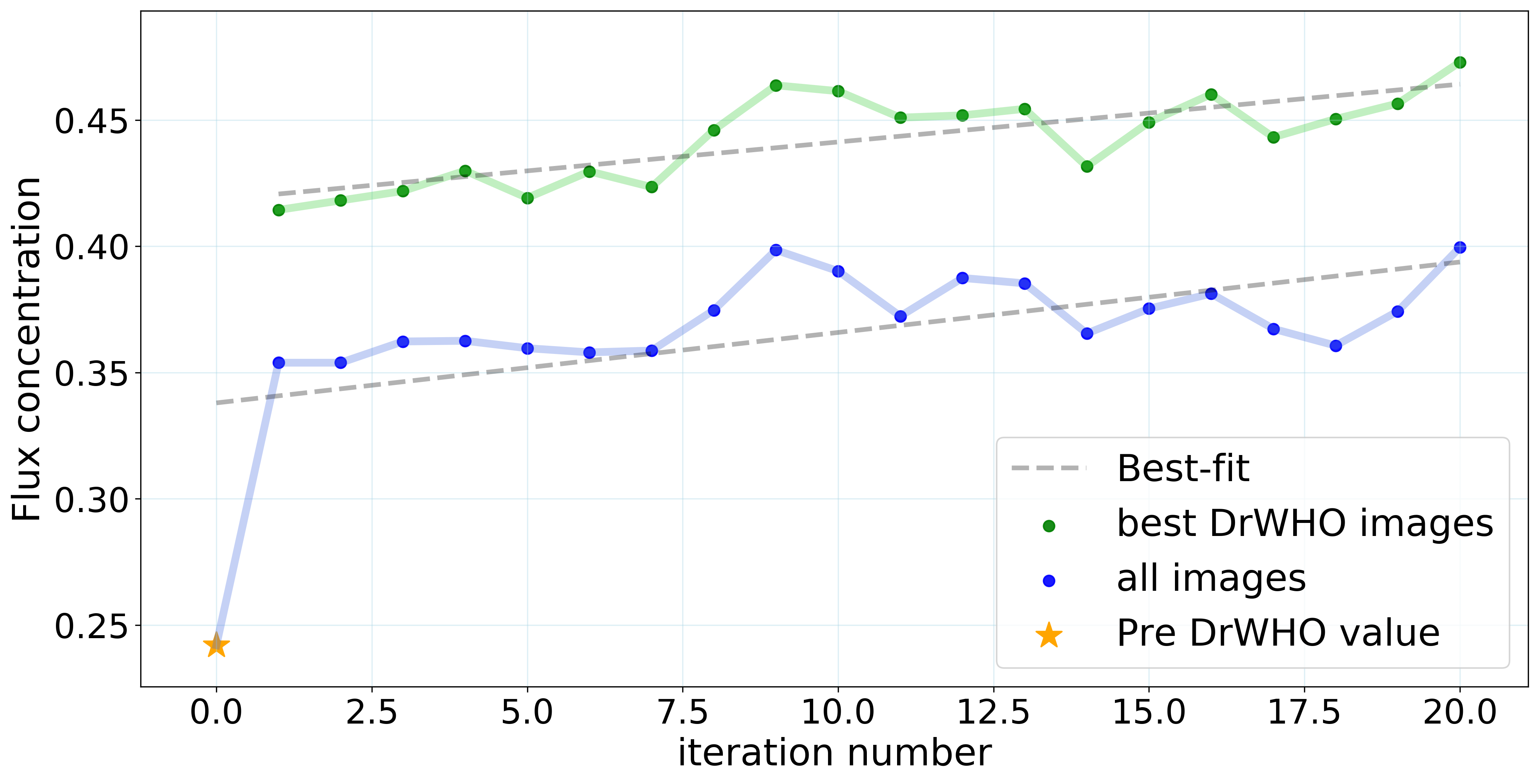}
\caption{Evolution of the FC over 20 iterations of DrWHO, corresponding to a period of 10 minutes, including the nFC of the PSF preceding the run. 
The blue curve corresponds to the evolution of the nFC of all the PSF averaged over the DrWHO iteration, while the green one corresponds to the evolution of the nFC of the 10\% PSF chosen by DrWHO. The best linear fit is presented. }
\label{fig:NNevolion}
\bigbreak
\includegraphics[width=.4\textwidth]{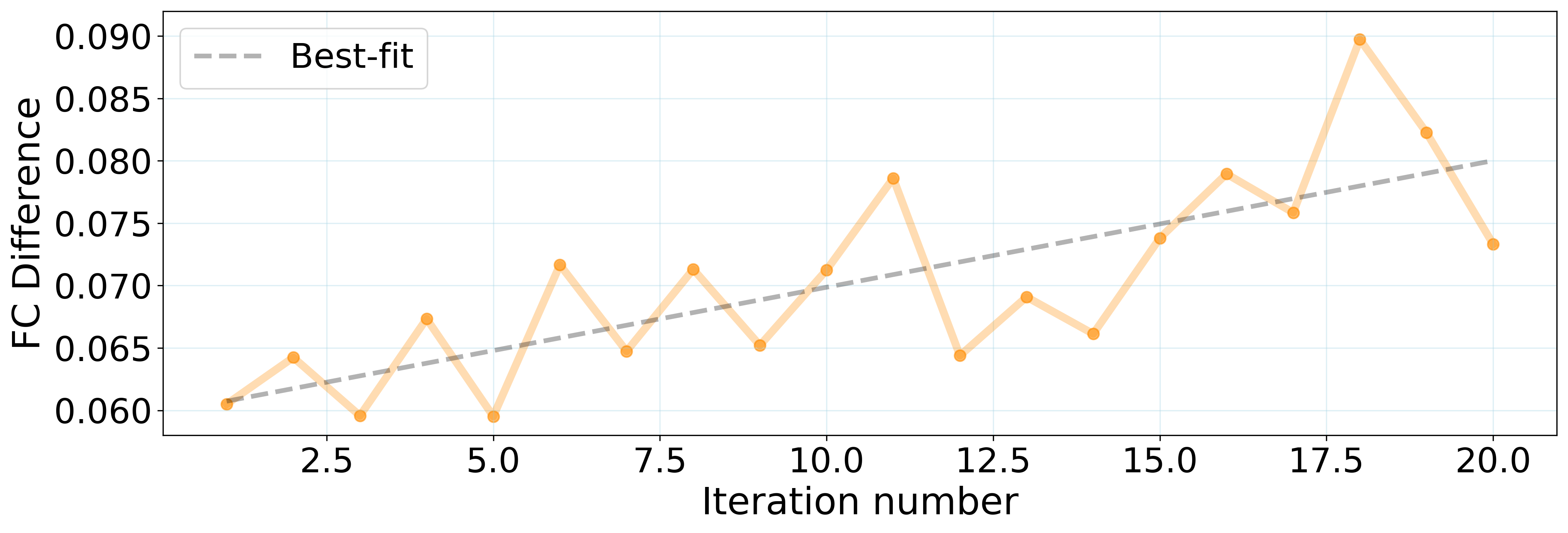}
\caption{Difference between the nFC of the mean of the best 10\% images, and the nFC of the mean of all the PSFs, over 20 iterations of DrWHO. We calculated the best linear fit to better understand the increase of the difference.}\label{fig:difference}
\end{figure}

\begin{figure}[t]
\centering
\includegraphics[width=.52\textwidth]{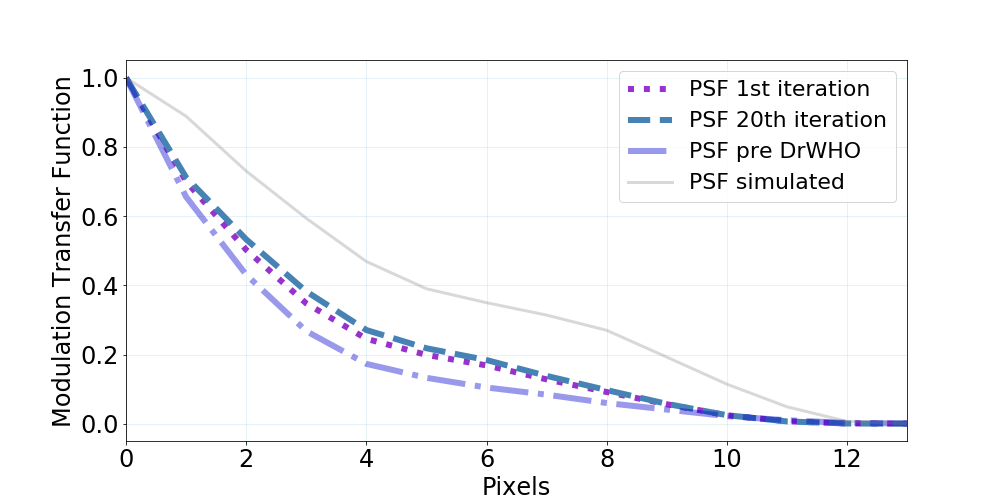}
\caption{MTF of the PSF before running DrWHO, of the 1st iteration, and the last iteration (cf Figure \ref{fig:SkyResult}), compared to the ideal simulated Vampires PSF.}
\label{fig:MTF}
\end{figure}

The comparison of the PSFs before, after the fist iteration, and after the last iteration of the DrWHO run is shown in Figure \ref{fig:SkyResult}. 
The evolution of the nFC over the 20 DrWHO iterations is shown in Figure \ref{fig:NNevolion}.
We plotted the evolution of the nFC for the 10\% best PSFs that were selected by the algorithm for the new reference computation, as well as the nFC for the average of all the PSFs for each iterations. The difference of those nFC is shown in Figure \ref{fig:difference}.\\
Furthermore, the Modulation Transfer Function (MTF) has been calculated for the PSFs before and after the DrWHO run, and is shown in Figure \ref{fig:MTF}. The MTF of the ideal simulated PSF has been added to the figure. This figure shows as well that most of the correction is done at the first iteration.

Here are the points highlighted in \citenum{DrWHO1}:
\begin{itemize}
    \item After 20 iterations, the PSF visually looks better and more circular, attesting a partial correction of the low orders;
    \item The major correction is applied at the first iteration, with the nFC jumping from 24.3\% to 35\%, comparably to simulations; 
    \item The FC keeps increasing throughout the run, as shown by the best linear fit of the two curves, which is something that was not observed in simulations, where the evolution was smoother and reached a plateau ;
    \item Considering PSFs averaged over each iteration, nFC increases from 28\%, to 35\% at the first iteration, to then 40\%: \textbf{the overall improvement is 15.7\%} ; 
    \item The best DrWHO images present a sharper increase than all the images together : Figure \ref{fig:difference} presents the difference of these two nFC over the algorithm run. Despite selecting only the best 10\% frames, DrWHO seems to be improving both the selected and unselected frames.
    We note that the algorithm has not fully converged yet and would have performed even better on a longer period of time ; 
    \item Finally, Figures \ref{fig:NNevolion} and \ref{fig:difference} show that even if the PSFs are subject to the variations in seeing over a few minutes, they overall get better in terms of the concentration of flux. Hence, DrWHO corrects the NCPA even if they are not as dominant as the dynamical wavefront residuals from atmospheric turbulence.
\end{itemize}

\subsubsection{WFS reference evolution}

\begin{figure}[t]
\centering
\includegraphics[width=.55\textwidth]{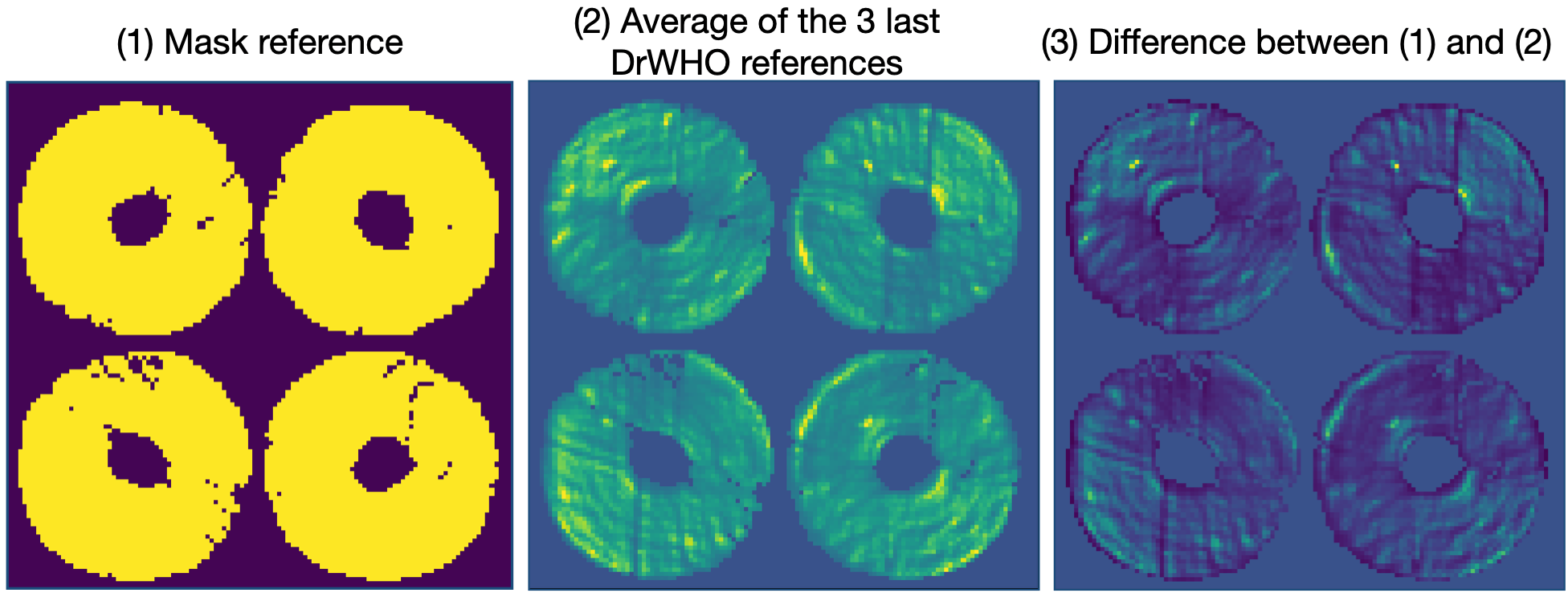}
\caption{Left: images of the PyWFS reference before the first iteration, referred to as \textit{Ref0}, set equal to the WFS mask. Middle: average of the last 3 iterations of DrWHO WFS references. The contrast scale for this image has been modified for better visualisation. Right: difference between the two images in the left and in the middle.}
\label{fig:reference_diff}
\end{figure}

\begin{figure}[t]
    \centering
    \includegraphics[width=.5\textwidth]{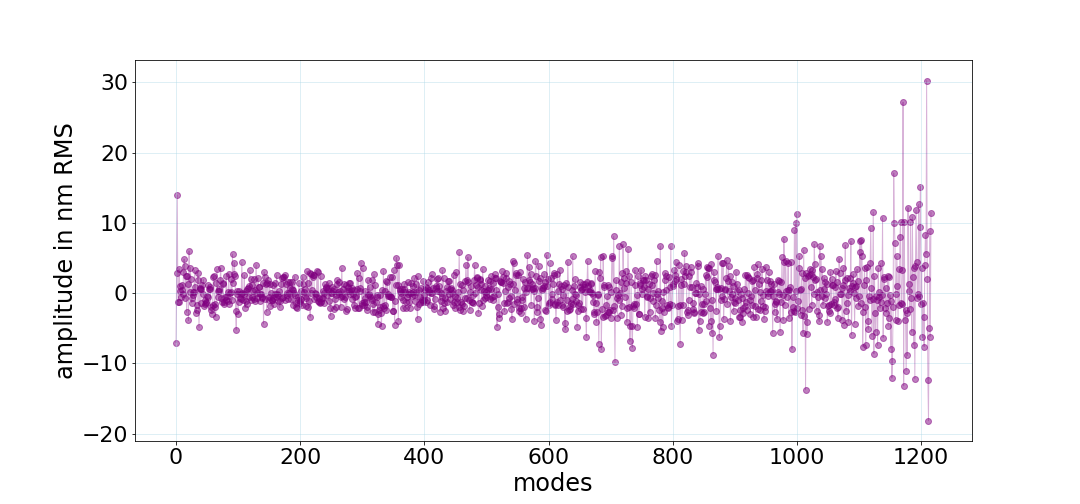}
    \caption{Modal decomposition of DrWHO correction, computed from the difference between the pre-DrWHO WFS reference and the average of the WFS references for the last 3 DrWHO iterations.}
    \label{fig:modes_proj}
    \bigbreak
    \includegraphics[width=.5\textwidth]{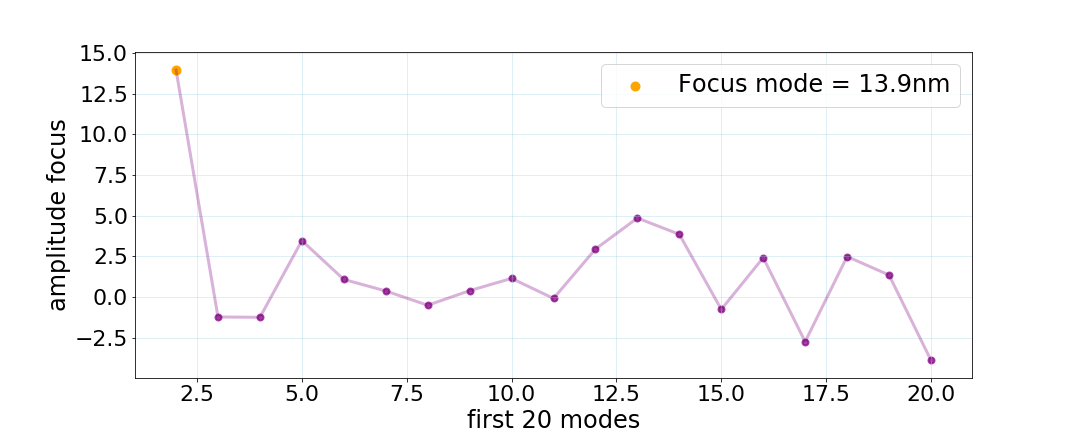}\\
    \includegraphics[width=.5\textwidth]{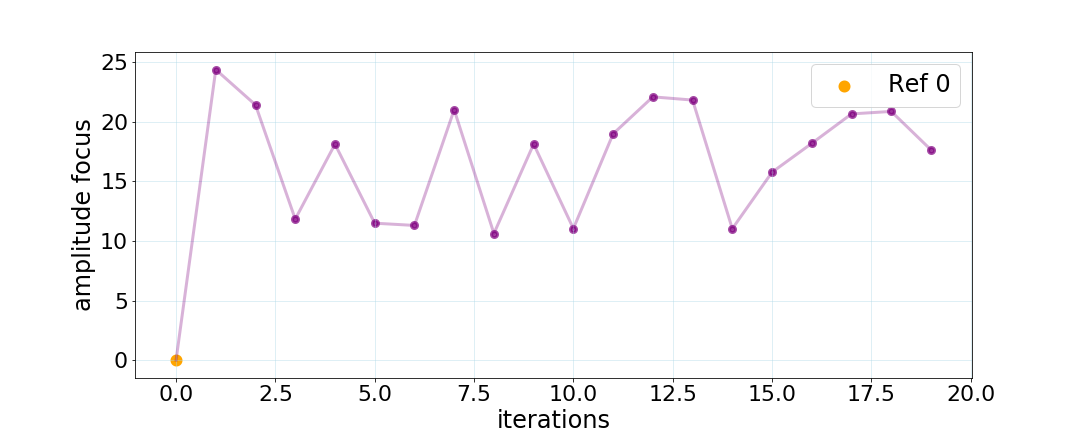}\\
    \includegraphics[width=.5\textwidth]{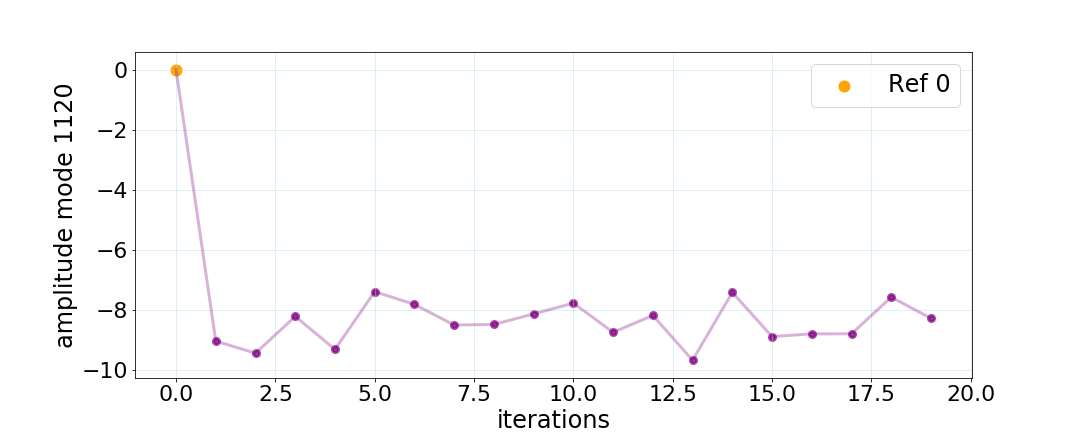}
    \caption{Top: modal decomposition of DrWHO correction for the first 20 modes, excluding tip-tilt. Focus is the major contributor with an amplitude of 13.9~nm RMS. 
    Middle: evolution of the focus WFS reference correction during the DrWHO run. 
    Bottom: evolution of high order mode 1120 WFS reference correction during the DrWHO run. }
    \label{fig:10_modes}
\end{figure}

Figure \ref{fig:reference_diff} compares the WFS reference prior to the DrWHO run, Ref 0, which is simply the WFS mask showing which pixels are active (i.e. respond to DM pokes) in the WFS image, and the average of the last 3 WFS references applied by DrWHO, during the on-sky 10-minute run. The averaging minimises the noise contribution.
The difference between the two references is also shown in Figure \ref{fig:reference_diff}.
This difference was then converted to wavefront mode coefficients by multiplication by the control matrix. Modal coefficients are shown in Figure \ref{fig:modes_proj} where the first three modes are tip, tilt, and focus and the following modes correspond to orthogonal modes, optimized for SCExAO's AO control law.

The total contribution of all modes is measured to be 116~nm RMS. However, as seen on Figure \ref{fig:modes_proj}, the high order modes, corresponding to the higher spatial frequencies, seem to be noise more than actual correction, which will be further explored thereafter. 

When calculating the total contribution of the modes by removing the modes beyond 900 (roughly where the modes amplitude starts increasing), we compute a correction of 70~nm RMS.

Figure \ref{fig:10_modes} (top) shows the first 20 wavefront mode values, excluding tip-tilt. Focus is the dominant contribution to the aberrations, with an amplitude of nearly 14~nm, while the other modes have values below 5~nm. We furthermore note that the higher modes from approximately the mode 900 to 1217, corresponding to higher frequencies, seem to have a considerable contribution in the overall correction. One would wonder whether those higher order contributions are just random noise, or meaningful behavior. We plotted the evolution of one example of a high mode (arbitrarily chosen for this paper to be the mode 1120), in the bottom of Figure \ref{fig:10_modes}. After the first iteration, the algorithm converges instantly towards a value of about -9~nm, with a standard deviation of approximately 2~nm RMS, which probably corresponds to a noise contribution. This observation is valid for the entirety of the high order modes. Table \ref{tab:mode_contribution} presents the contribution of the modes in nm RMS, in the following cases: in considering all the modes except tip-tilt (from 2 to 1217) ; then in removing the higher order modes (from 2 to 900 only) to show what the contribution would be without the higher spatial frequencies ; then the low order modes (from 2 to 20), and finally the focus contribution, which is  13.9~nm.

\begin{table}[ht]
\begin{tabular}{c|l|l|l|l}
\textbf{Modes}                                                            & 2 to 1217 & 2 to 900 & 2 to 20 & \multicolumn{1}{c}{Focus} \\ \hline
\textbf{\begin{tabular}[c]{@{}c@{}}Contribution \\ (nm RMS)\end{tabular}} & 116.3     & 70.3     & 17.2    & 13.9   
\end{tabular}
\caption{Contribution of the modes of the difference of Figure \ref{fig:difference} in nm RMS}
\label{tab:mode_contribution}
\end{table}

\section{On-sky preliminary coronagraphic tests}
\label{sec:coro_tests}
After successful tests on non-coronagraphic images, DrWHO has been adjusted to fit coronagraphic images requirements. The only difference of the algorithm lies in the selection criteria for the images, from the maximization of the FC on PSF, to the minimization of the intensity on coronagraphic images, in order to optimize the contrast. The fast focal plane camera was the science viewing camera, running on the IR on H band, which is an InGaAs array. \\
We present here some preliminary results, one should however keep in mind that tests are still ongoing. 

\subsection{Observation setup}

DrWHO was tested on-sky on coronagraphic images during the engineering night of July 14 UT, 2021. The successful test was of 5 iterations, each of 30 seconds. 
1200 modes were corrected. The observation conditions at that time were a seeing of 1.2", and a wind speed of 40 m/s. The observed target was the star Vega.
The window size of the image for the selection of the quality criteria was of 50x50 pixels, corresponding to 0.81x0.81 arcsec field of view.

\subsubsection{Image quality evolution}

The goal is here to optimize the contrast of the image. Hence, the selection criteria for DrWHO becomes the minimization of the intensity, which is the metric we are focusing on. \\

\begin{figure*}[t]
\begin{center}
\centering
\includegraphics[width=0.7\textwidth]{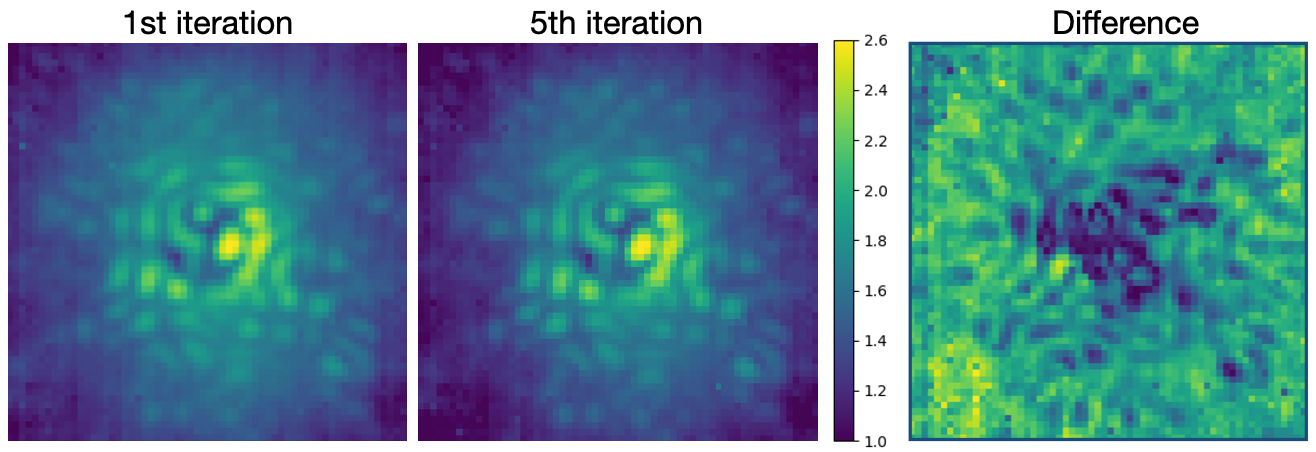}
\caption{Left and middle: comparison of the coronagraphic image before and after DrWHO, in logarithmic scale. Right: difference of those two images, normalized by the pre-DrWHO image. The dark values are negative and indicate improvement.}
\label{fig:coro_results}
\end{center}
\end{figure*}

\begin{figure*}[t]
\begin{center}
\centering
\includegraphics[width=0.8\textwidth]{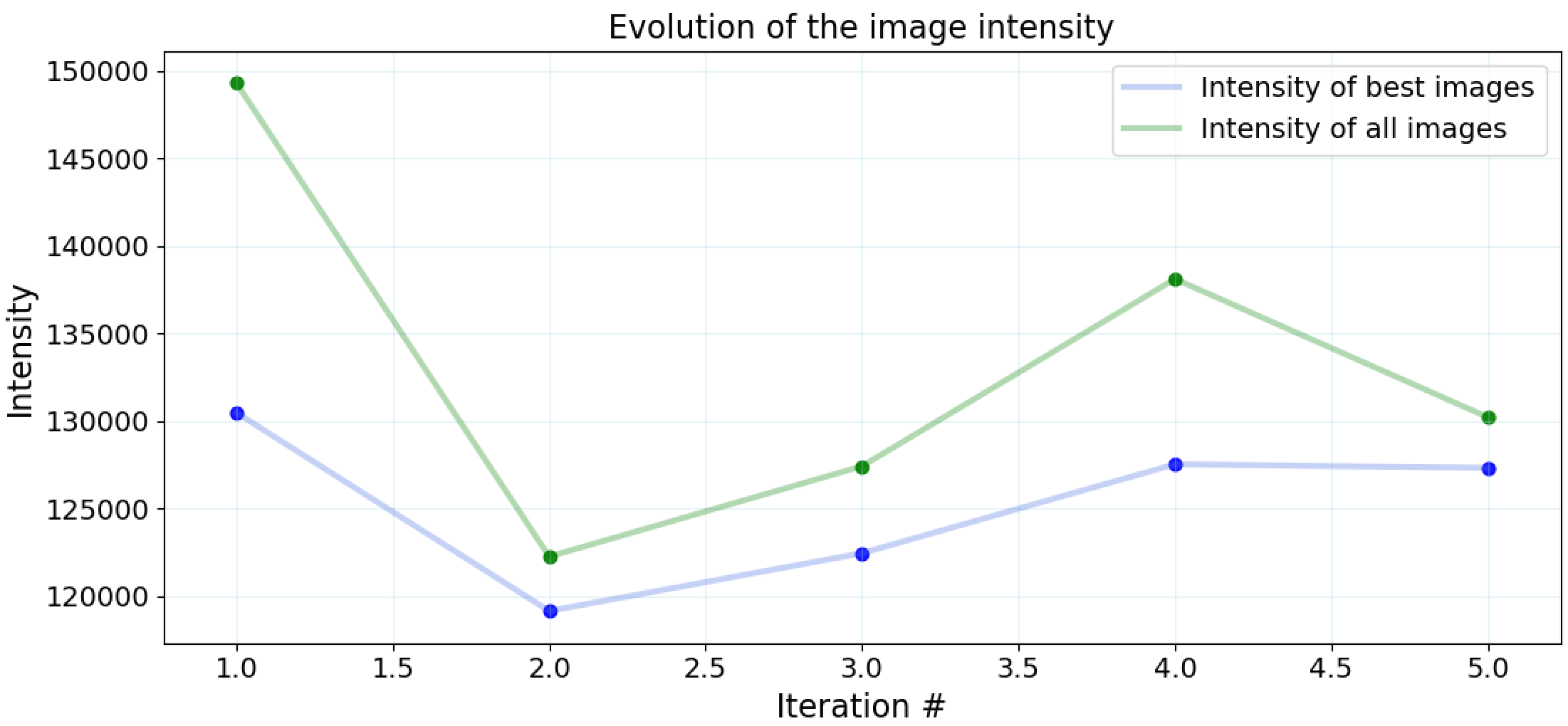}
\caption{Evolution of the intensity for each iteration of the on-sky test on coronagraphic images.}
\label{fig:intensity_coro}
\end{center}
\end{figure*}

The comparison of the coronagraphic images before and after the DrWHO run is shown in Figure \ref{fig:coro_results}, left and middle. The right part presents the difference of the images before and after the run, normalized by the pre-DrWHO image. This shows that the center of the image is where most of the correction is made, in other words the essential of the correction is done over the lower order modes, which is the main contribution of the NCPA. The edges of this figure correspond to the higher order modes, highlighting the high seeing during this test. This is indicative of a gain, however further tests are needed to quantify it. 
The evolution of the intensity over the 5 DrWHO iterations is shown in Figure \ref{fig:intensity_coro}. We plotted the evolution of the intensity for the 10\% best coronagraphic images that were selected by the algorithm for the new reference computation, as well as the intensity for the average of all the coronagraphic images for each iterations.

A few points are worth noting: 
\begin{itemize}
    \item First, the observations conditions were considerably worse compared to the DrWHO test on non-coronagraphic images;
    \item Figure \ref{fig:coro_results} shows that the coronagraphic images after the 5 iterations, the contrast is better. This is confirmed by the right image of the Figure, highlighting the correction of the low orders;
    \item There is an improvement of the image contrast, as shown in Figure \ref{fig:intensity_coro}, from the first iteration of the algorithm. This improvement is however relatively minor, which is, at least partially, the result of the high seeing during the observations;  
\end{itemize}

\section{Discussion}\label{sec:Discussion}
Below are some noticeable characteristics of the DrWHO algorithm that have been demonstrated in this proceeding: 
\begin{enumerate}
    \item DrWHO is robust, for both non-coronagraphic and coronagraphic use: the algorithm slowly but surely converges towards a better compensation of static and slow aberrations. As it is based on lucky imaging, it relies on some realisations of residual atmospheric turbulence to reduce wavefront aberrations. DrWHO contributes to getting closer to an absolute sensor, driving the WFS reference to optimize images in the focal plane. 
    \item It does not rely on any model, and does not rely on using the DM for probing, making it compatible with other wavefront control techniques: DrWHO's approach is different from active cancellation speckle, because it simply does a statistical selection using the natural dynamical atmospheric residuals for probing, instead of adding artificial probes.
    \item It does not require modification of the AO loop control scheme other than WFS reference updates.
    \item It does not make any linear assumption, so it is not concerned with non-linearities of the system, such as WFS non-linearity.
    \item It is flexible in the parameters, whether it is the score (Strehl, contrast, intensity, etc.), or the selection fraction and the time of the iteration, making it adaptable to different weather conditions and AO systems with different characteristics and scientific goals.
    \item DrWHO requires decent seeing for being efficient, otherwise the atmospheric residuals will simply overcome the NCPA. However, tests on seeing at 1.2" show DrWHO does not worsen the image quality, and still improves it. 
    \item However, DrWHO needs a fast focal plane camera as close as possible to the science camera, to more accurately compensate the NCPA and low aberrations. 
\end{enumerate}

There are numerous ways to improve the algorithm efficiency and convergence speed. For example, a possible extension is to parallelize DrWHO in terms of spatial frequencies. If there is a good understanding of the relationship between the input wavefront and the focal plane, as for example a Fourier relationship, DrWHO can optimize separate spatial frequencies independently.

\section{Conclusion}

The DrWHO algorithm has first been tested on non-coronagraphic images in the visible, by being initially implemented on an ExAO simulator, then on the SCExAO instrument and tested on-sky, at the Subaru Telescope. Secondly, DrWHO was validated on coronagraphic images in the IR. 

Results presented here demonstrate its ability to measure and compensate for NCPA, considered to be a limiting factor in the detection and characterisation of exoplanets in high-contrast imaging observations. To quantify the quality of the PSF, we used the Strehl ratio for the simulations, and the \textit{Flux Concentration} for the SCExAO run. For coronagraphic images, we used the total intensity of the image, which is correlated with the image contrast. 
For the simulation, we observed a nearly perfect compensation of the added NCPA, at the first iteration.

For the on-sky test, we combined the PyWFS measurements and the focal plane images in visible from the VAMPIRES module for non-coronagraphic tests, and from the fast focal plane IR camera for coronagraphic tests. 
For non-coronagraphic tests, when DrWHO was running, the PSF significantly improved (15.7\% relative improvement in flux concentration) over 20 iterations of 30 seconds, for a total of 10 minutes. The best PSF have a sharper increase over the run of the algorithm, reaching 48\% of the simulated ideal PSF flux concentration. 
The visible improvement in the image quality was confirmed with the calculation of the MTF before and after the DrWHO run.

Regarding coronagraphic tests, the preliminary results in relatively bad seeing indicate that DrWHO does improve the image contrast, and the corrected modes are mainly the low orders. Further tests are ongoing to fully quantify the extend of the potential of the algorithm.\\

Overall, we show that DrWHO is able to improve the wavefront quality arriving on the science camera, and partially correct for the static and quasi-static NCPA, over a relatively short period of time of a few seconds. We show that the algorithm converges rapidly, with about half of the correction achieved by the first iteration. This has been observed in simulation and on-sky.
Furthermore, the correction of the NCPA is effective even if it is considerably smaller than the dynamic of the atmosphere.

The characteristics of DrWHO are enumerated in section \ref{sec:Discussion}. The main strong points are the robustness of the algorithm, its independence from any model, making it compatible with other wavefront control methods, and that it does not rely on any linear assumption, thus particularly fit for a PyWFS. DrWHO combines focal plane wavefront sensing and the WFS. \\

A possible extension of DrWHO is to parallelize it in terms of spatial frequencies, which would allow to correct speckles individually and to observe fainter targets. 

This paper is the first proof of concept of the algorithm on coronagraphic images, following \cite{DrWHO1}, and presents some first results of its image quality and stability enhancement. 
DrWHO shows a wide potential of improvement, which will be presented in future work.



\acknowledgments 
 
Based [in part] on data collected at Subaru Telescope, which is operated by the National Astronomical Observatory of Japan. The development of SCExAO was supported by the Japan Society for the Promotion of Science (Grant-in-Aid for Research \#23340051, \#26220704, \#23103002, \#19H00703 \& \#19H00695), the Astrobiology Center of the National Institutes of Natural Sciences, Japan, the Mt Cuba Foundation and the director's contingency fund at Subaru Telescope. The authors wish to recognize and acknowledge the very significant cultural role and reverence that the summit of Maunakea has always had within the Hawaiian community. We are most fortunate to have the opportunity to conduct observations from this mountain. NS acknowledges support from the PSL Iris-OCAV project. NS and VD acknowledge support from NASA (Grant \#80NSSC19K0336). KA acknowledges support from the Heising-Simons foundation. 

\bibliography{report} 
\bibliographystyle{lyotspiebib} 

\end{document}